# Arp@VST: Legacy Survey of the Arp Peculiar Galaxies with the VST

Marilena Spavone[1]
Chiara Buttitta[1]
Rosa Calvi[1]
Alessandro Loni[1]
and the VST Team

[1] INAF–Astronomical Observatory of Capodimonte, Naples, Italy

Arp@VST is a public observing programme, conducted at the VLT Survey Telescope (VST) hosted at ESO's Paranal Observatory. It aims to revisit the Arp catalogue by creating a public survey. The Atlas of Peculiar Galaxies was produced by Halton Arp in 1966 and contains 338 galaxies with distorted morphologies and/or interacting systems. Given the excellent capabilities of the VST to map the galaxies' structure down to low surface brightness levels, for this project we will acquire deep, multi-band (g, r, i, Hα) images for all the Arp galaxies visible from Paranal Observatory (Declination < +10 deg). Being a public survey, the reduced data will be released via the ESO Science Archive as soon as they are processed.

## Overview of the Arp catalogue

The images for Arp's Atlas of Peculiar Galaxies (Arp, 1966) were acquired with the Hale and Schmidt telescopes of the Carnegie Observatories, providing the best photographic plates possible at the time, and they revealed faint galactic details at very high resolution.

As stated in the preface of the Arp catalogue, "the peculiarities of the galaxies pictured in this Atlas represent perturbations, deformations, and interactions which should enable us to analyse the nature of the real galaxies which we observe and which are too remote to experiment on directly." This sentence summarises Arp's main goal for the catalogue. At that time, the physical processes that shape the galaxies in the Hubble sequence were not fully understood. The Arp catalogue therefore aimed to provide examples of the different kinds of peculiar structures found in galaxies, to be considered as 'laboratories' where astronomers can study the mechanisms acting on main sequence galaxies, i.e. spiral or elliptical galaxies.

The atlas does not include all kinds of peculiar galaxies known to date. It collects several examples of gravitational interactions between galaxies, grouped as follows:
– individual peculiar spiral galaxies or spiral galaxies that apparently have small companions;
– elliptical-like galaxies;
– individual or groups of galaxies with distorted morphology;
– galaxy pairs.

Nowadays, the physical processes that lead to the formation of galaxies and, on a larger scale, of galaxy clusters seem to be well described by the Lambda-Cold Dark Matter (LCDM) model. In this framework, clusters of galaxies are expected to grow over time, accreting smaller groups along filaments, driven by the effect of gravity generated by the total matter content (for example, White & Rees, 1978; Bullock et al., 2001). In the deep potential well at the cluster centre, galaxies continue to undergo active mass assembly. In this process, gravitational interactions and merging between systems of comparable mass and/or smaller objects play a fundamental role in defining their morphology and kinematics (Toomre & Toomre, 1972). As a result of the repeated accretion events, the galaxies' outskirts are populated by stellar streams, shells, or tidal tails.

On the observational side, owing to the diffuse and extremely faint nature of these features their detection and analysis are the most challenging tasks. In the last two decades, the long integration times and incrementally larger covered areas of current facilities have allowed focused, deep, multi-band imaging and spectroscopic surveys to deliver data with the depth ($\mu_g$ ~ 28–31 mag arcsec$^{-2}$) and resolution to reach unprecedented limits in the galaxy outskirts and the intra-cluster space. This has enabled extensive analyses of the light and colour distributions, kinematics and stellar populations of interacting and peculiar systems (see, for example, Duc et al., 2015; Iodice et al., 2016; Mihos et al., 2017; Spavone et al., 2020; Danieli et al., 2020; Trujillo et al., 2021; Miller et al., 2021; Spavone et al., 2022; Gilhuly et al., 2022, and references therein).

## The motivation for Arp@VST

Given the excellent capabilities of the VST to map the galaxies' outskirts down to the low surface brightness (LSB) levels needed to detect and study the remnants of gravitational interactions and merging (see Peletier et al., 2020 and Iodice et al., 2021 for reviews), the Arp@VST[1] project will use the VST to provide deep multi-band (g, r, i, Hα) images of the peculiar galaxies in the Arp catalogue that are visible from Paranal Observatory. This is a public survey, since the reduced data will soon be made available to the astronomical community. As described in detail in the next sections, the new Arp@VST catalogue will have a twofold value. It might be a treasure trove for astrophysical studies of galaxy interactions, the formation of tidal dwarf galaxies, and star-forming regions. And in addition, the wonderful colour-composite images that can be obtained and released through the media channels have a strong public engagement potential to attract the attention of young students towards astrophysics.

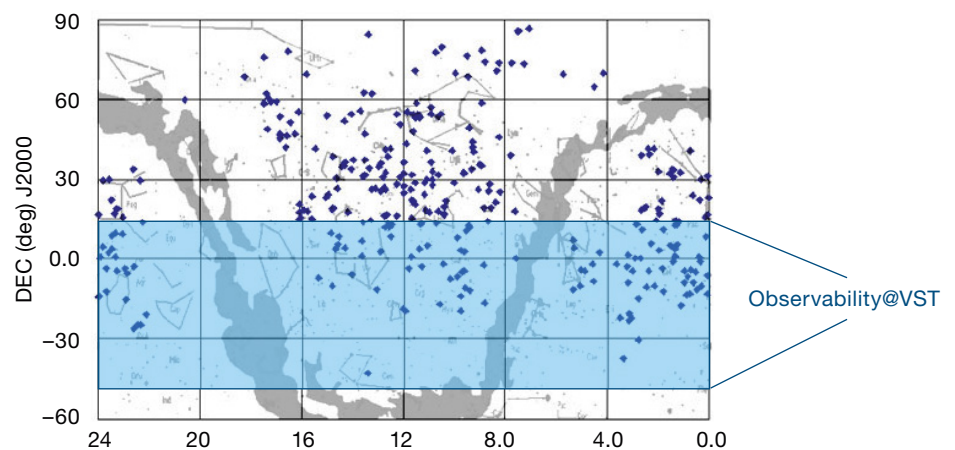

Figure 1. Location of the Arp peculiar galaxies. The blue shaded area indicates the region of the targets' observability from the VST at ESO's Paranal Observatory.

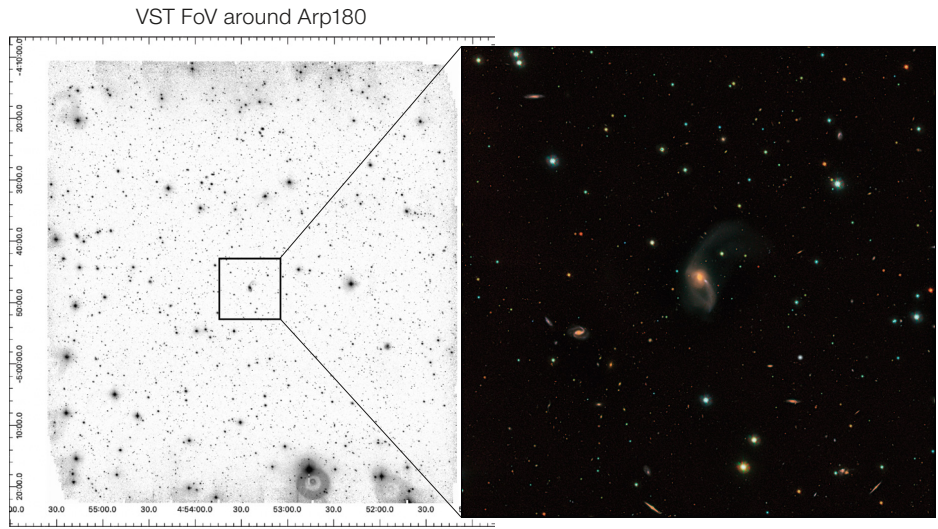

Figure 2. VST images of Arp180. Left panel: entire OmegaCAM field of view (1 deg$^2$) centred on the target, in the $g$ band. Right panel: colour-composite ($g$, $r$, $i$, H$\alpha$) image of an enlarged region around Arp180.

Survey plan and science deliverables

We plan to acquire deep images (2.5 hours of integration time) in the $g$ and $r$ bands, down to $\mu_g \sim 28$–31 mag arcsec$^{-2}$, for all the Arp galaxies visible from Paranal Observatory), to detect the LSB features (in the form of stellar streams or tidal tails) out to unprecedented distances from the centre of the target (around 10 $R_e$, see Figure 2.). Shallower images (integration time about 1 hour) will be acquired in the $i$ band and with the H$\alpha$ filter. The full list of targets[2] is made up of more than 120 objects.

There are two main reasons for using the VST for this project: the large field of view (FOV) and the high arcsec-level angular resolution of OmegaCam. The large FOV is necessary to map the structure of the galaxies' outskirts out. Eighteen of the Arp galaxies have been observed with the Hubble Space Telescope (HST). Whereas HST images have better angular resolution than ground-based VST images, the FOV of the HST is significantly smaller than 1 deg$^2$. Specifically, Arp images taken with the HST's ACS/WFC instruments have a FOV approximately 300 times smaller than that of the VST, while those taken with the WFC3 and WFPC2 instruments are about 500 times smaller. They therefore provide only a limited view of the faint features in the outskirts of the Arp target galaxies. With the large area covered by the VST, we can extend the detection and analysis of these features. In addition, the large FOV, combined with the resolution of 0.21 arcsec pixel$^{-1}$, is fundamental to improving the detection of the satellite galaxies in the outskirts and to resolving the fine features in the tidal tails, such as the star-forming clumps.

This survey aims to gather valuable insights into the fascinating and peculiar structures of Arp galaxies, enhancing our understanding of their formation and evolution. To this end, the main science deliverables that this project can offer are presented below.

Interaction and merging: combining the deep images acquired in the $g$ and $r$ bands and the large area covered by the VST data, tidal tails or stellar streams can be detected down to the faintest levels of $\mu_g \sim 28$–31 mag arcsec$^{-2}$ and out to unexplored regions at larger radii from the target centre. The structure of these features can be directly compared with the predictions of all phases of the gravitational merging of galaxies with different mass ratios. The main outcome of these studies is to provide more stringent constraints on the formation of the massive early-type galaxies and the mass assembly process.

Detection of small satellites: to date, the confirmation of the LCDM cosmological model relies on our ability to find the baryonic counterparts of the predicted low-mass dark matter halos, which means deriving a complete census of the faintest, less massive stellar systems such as dwarf galaxies. The depth and the large FOV of the acquired images can offer a unique opportunity to improve the detection of the satellite dwarf galaxies, in particular those at the lowest surface brightness and diffuse systems.

Detection and formation of the tidal dwarfs and tidal ultra difuse galaxies: as a follow-up to the detection of dwarf galaxies mentioned above, this project provides an excellent sample of interacting galaxies, in which a statistically significant sample of tidal-dwarf galaxies can be detected. These kinds of systems, formed in the tidal tails of merging spiral galaxies, are particularly interesting since they are dark-matter-free (Duc et al., 2015). The structure and stability of these galaxies are still an open issue to be investigated.

Star formation regions: the H$\alpha$ images acquired with this project can be used to identify and study the star-forming regions and determine whether they might have been affected by environmental interactions. Any asymmetry in the star formation rate map derived from H$\alpha$ can indicate whether (and where) a galaxy is experiencing a star formation rate enhancement or a possible quenching of star formation. In addition, the displacement of ionised gas with respect to the stellar component can be identified. This might reveal ongoing (and possibly subtle) hydrodynamical interactions. The results will offer an exquisite opportunity to identify interesting regions for further follow-up to understand how gas is being removed.

Figure 3. Colour-composite (*g*, *r*, *i*, Hα) VST images of an enlarged region around Arp 251 (upper panel) and Arp 257 (lower panel).

**Spectral Energy Distribution:** by combining the multi-band images released by Arp@VST, which cover the optical wavelength range, with available data in other bands, from ultraviolet to near-infrared, the spectral energy distribution can be addressed for each target of the sample.

## Arp galaxies with the VST

To date, during the first six months or so of this project, we have acquired data for around 10 targets[2]. Reduced data for four targets have already been released to the community, and they are available in the ESO Science Archive[3].

Figures 2 and 3 show the colour-composite VST images for three fascinating objects in the catalogue, described below.

**Arp 180** – This is a pair of interacting galaxies in an advanced stage of merging (Figure 2), located at a distance of approximately 60 Mpc from us. The nuclei of the two progenitor galaxies are no longer distinguishable, and a bright, roundish central object is forming. Two very extended tidal tails (measuring approximately 25 kpc to the north and 15 kpc to the south) dominate the scene: one to the south, resembling a closed loop, and another to the north, forming an arm-like structure. Near the centre of the system, a bright, southwest blue knot is visible, likely tracing ongoing star formation.

**Arp 251** – This is a very distant group of three spiral galaxies, located approximately 300 Mpc away, showing clear signs of gravitational interaction (Figure 3, upper panel). The bluish colours and clumpy spiral arms in all three systems suggest strong star formation activity. The spiral arms of the northern galaxy and the one to the west appear to be significantly affected by the ongoing interaction, resulting in a more unfolded morphology compared to typical, unperturbed spiral galaxies. Two additional interesting objects are also visible in the image: to the northwest, a foreground warped spiral galaxy (LEDA 173089) at a distance of around 200 Mpc, and to the east, a

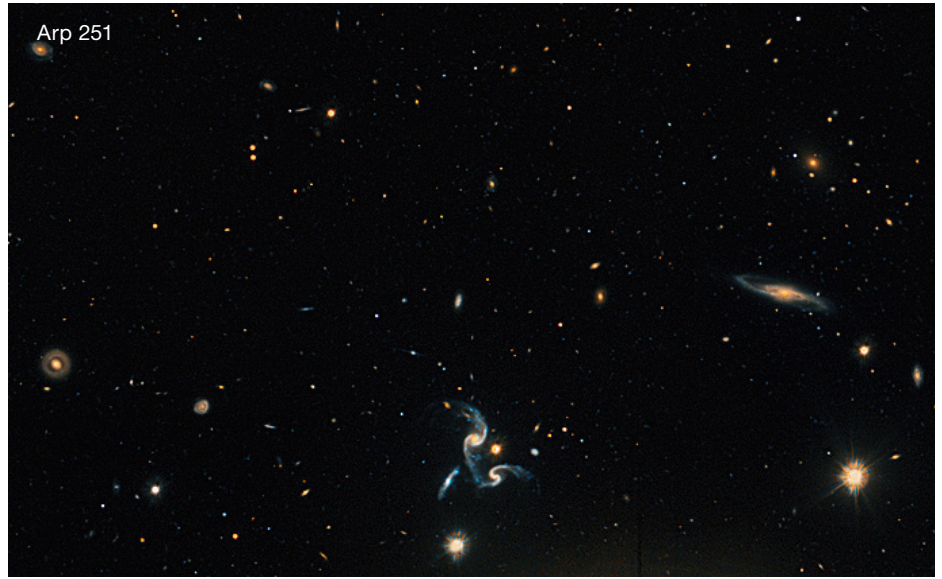

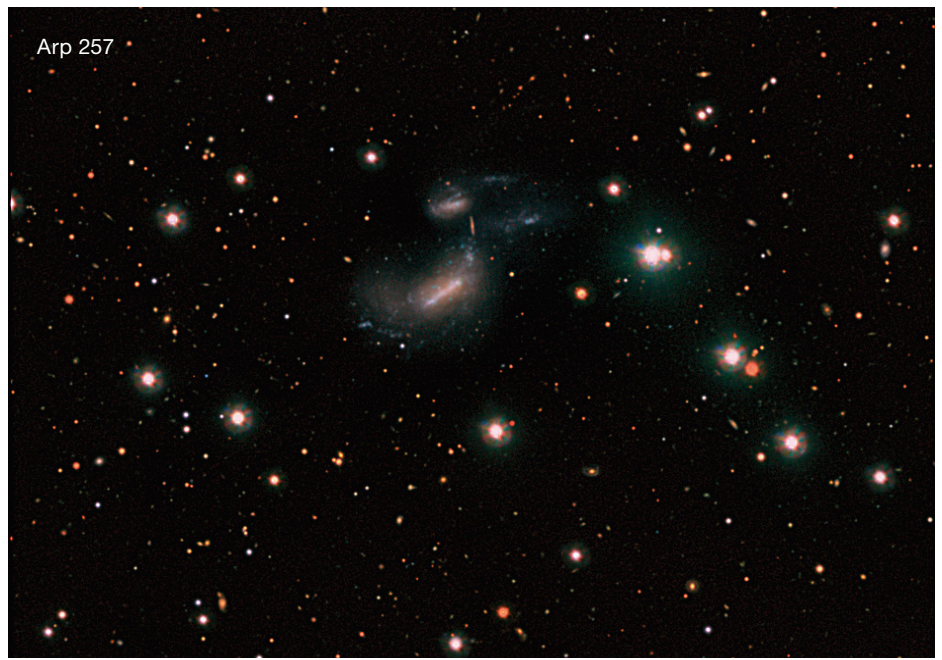

Caldwell-like spiral galaxy (LEDA 932482) with an unknown redshift.

**Arp 257** – This is an intriguing nearby (roughly 43 Mpc) pair of interacting galaxies with distorted spiral features and a thin tidal arc linking them (Figure 3, lower panel). The system was first noted by Arp, who described it as "galaxies joined by a segment of thin arc". Both galaxies exhibit clear signs of mutual interaction and distortion, along with prominent blue star-forming regions — evidence that their encounter is stirring interstellar gas and triggering new star formation. Like Arp 251, this system is a clear example of an ongoing merger, where tidal bridges and clumps dominate the morphology.

## Data sharing

Arp@VST is a public observing programme that is committed to producing and publicly sharing the fully reduced dataset. We therefore encourage scientists from any country and affiliation to request a specific dataset (either already observed or planned for observation) to develop a science project, including research for a Master's or PhD thesis. The VST team is responsible for preparing observation blocks (OBs), scheduling observations, and reducing the data for release. For further details, please read the instructions provided[4].

Acknowledgements

The authors acknowledge financial support from INAF to the VST Coordination Centre[5].

Links

[1] Arp@VST website: https://sites.google.com/inaf.it/arp-vst/home-page?authuser=3
[2] Arp@VST targets: https://sites.google.com/inaf.it/arp-vst/targets
[3] Arp@VST data in the ESO Science Archive: https://doi.eso.org/10.18727/archive/99
[4] Arp@VST data request: https://sites.google.com/inaf.it/arp-vst/data-request
[5] VST Coordination Centre Website: https://vst.inaf.it/home

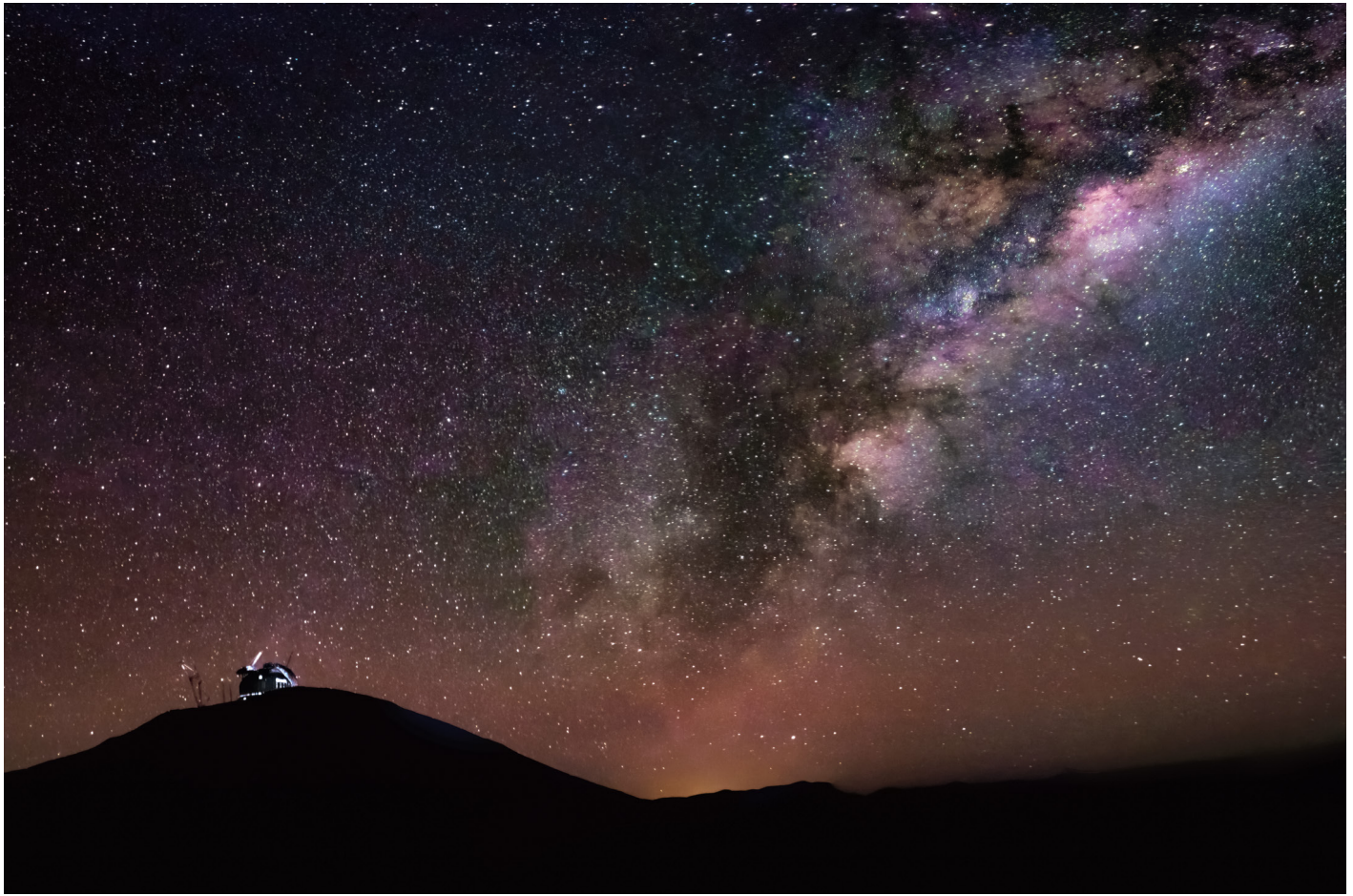

This image shows ESO's Extremely Large Telescope (ELT) underneath a gorgeous starry night sky, perched atop the mountain of Cerro Armazones, Chile, like a ship cresting a great, shadowy wave.

It's a view of the telescope we don't often see, taken from the Rolf Chini Cerro Murphy Observatory, a nearby facility hosted within ESO's Paranal Observatory. However, if you look closely, you might see the familiar sight of cranes surrounding the ELT's construction site, one of them appearing as an unusual thin white line above the illuminated telescope's dome.

The ELT is set for completion at the end of the decade, when it will begin to search the sea of stars seen here — one of the most pristine skies anywhere on Earth, for which the Atacama Desert has a rich and important heritage.